\documentclass[numberedappendix,apj]{emulateapj}
\usepackage{amsmath,color}

\newcommand{\bl}{}
\shorttitle{Measuring Turbulent Diffusion}
\shortauthors{Owen, J. E.}

\begin{document}


\title{Snow-lines as probes of turbulent diffusion in protoplanetary discs}


\author{James E. Owen}
\affil{Canadian Institute for Theoretical Astrophysics, 60 St George Street, Toronto, M5S 3H8, ON, CANADA}

%\author{C. D. Biemesderfer\altaffilmark{4,5}}
%\affil{National Optical Astronomy Observatories, Tucson, AZ 85719}
%\email{aastex-help@aas.org}
%
%\and
%
%\author{R. J. Hanisch\altaffilmark{5}}
%\affil{Space Telescope Science Institute, Baltimore, MD 21218}


%\altaffiltext{1}{Visiting Astronomer, Cerro Tololo Inter-American Observatory.
%CTIO is operated by AURA, Inc.\ under contract to the National Science
%\altaffiltext{2}{Society of Fellows, Harvard University.}
%\altaffiltext{3}{present address: Center for Astrophysics,
%    60 Garden Street, Cambridge, MA 02138}
%\altaffiltext{4}{Visiting Programmer, Space Telescope Science Institute}
%\altaffiltext{5}{Patron, Alonso's Bar and Grill}


\begin{abstract}
Sharp chemical discontinuities can occur in protoplanetary discs, particularly at `snow-lines' where a gas-phase species freezes out to form ice grains. Such sharp discontinuities will diffuse out due to the turbulence suspected to drive angular momentum transport in accretion discs. We demonstrate that the concentration gradient  - in the vicinity of the snow-line - of a species present outside a snow-line but destroyed inside is strongly sensitive to the level of turbulent diffusion (provided the chemical and transport time-scales are decoupled) {\bl and provides a direct measurement of the radial `Schmidt number' (the ratio of the angular momentum transport to radial turbulent diffusion)}. Taking as an example the tracer species N$_2$H$^+$, which is expected to be destroyed inside the CO snow-line (as recently observed in TW Hya) we show that {\it ALMA} observations possess significant angular resolution to constrain the Schmidt number. {\bl Since different turbulent driving mechanisms predict different Schmidt numbers, a direct measurement of the Schmidt number in accretion discs would allow inferences about the nature of the turbulence to be made}.     

\end{abstract}





\section{Introduction}

Astrophysical discs are observed to transport angular momentum. It has been  hypothesised that such discs are transporting angular momentum through a turbulent process \citep{shakura73}.  Despite decades of theoretical research we still lack a sufficient understanding of turbulence to make quantitative predictions. Several candidate processes exist to sustain turbulence in these discs: the magneto-rotational instability (MRI) \citep[e.g.][]{balbus91}, gravitational instability (e.g. Lodato \& Rice 2004, 2005) the vertical shear instability \citep{nelson13} and the baroclinc instability \citep[e.g.][]{klahr03}. The MRI is still the leading candidate for angular momentum transport in accretion discs (see \citealt{turner14} for a recent review). However, at the low temperatures and ionization fractions expected in protoplanetary discs, non-ideal MHD effects become important, qualitatively changing the nature of the turbulence and its associated transport properties, or rendering it ineffective resulting in `dead-zones' \citep[e.g.][]{gammie96}. 

A corollary to the angular momentum transport problem in protoplanetary discs is the transport of dust particles. %Such dust particles are the building blocks for planetesimals and planets, therefore, it is crucial to understand their dynamics. 
There is large amounts of astrophysical \citep[e.g][]{bouwman01,dullemond06,hughes10,owen_edge_on} and cosmochemical \citep[e.g][]{gail01,bockel02,jacquet12,jacquet13} evidence to suggest large scale radial transport  of dust particles occurs. %and is an important process. 
In particular, the level of diffusion that results from the turbulence is an unknown parameter. The turbulent diffusion coefficient ($D_g$) is often parametrised in terms of the turbulent kinematic viscosity ($\nu$) as $D_g=\nu/Sc$ where $Sc$ is the Schmidt number. Assumptions of isotropic, Kolmogorov-like turbulence lead to the inference that $Sc\approx 1$ \citep{youdin07}; however, a large range of reported experimental/theoretical/numerical values exist that span approximately two orders of magnitude in the range $Sc=0.1-10$ \citep{prinn90,dubrulle91,lathrop92,carballido05,johansen06,youdin07,zhu14}. %Furthermore, it cannot be  assumed {\it a priori} that the Schmidt number is isotropic; in-fact recent simulations predict the Schmidt number is different in the radial and vertical direction and depends on location \citep{zhu14}.

Direct observational measurements of the strength of turbulent diffusion in the continuum will be complicated by dust-drag, grain growth/fragmentation and optical depth effects. However, it is not only dust particles that will experience turbulent diffusion; any rare, gas-phase species (tracer species) where changes in temperature results in a concentration gradient will also experience turbulent diffusion. If, for example, a tracer species is predominately produced at a given radius,  the tracer will then diffuse away from this radius, with a distribution that is strongly dependent on the value of the Schmidt number \citep{clarke88}. \citet{jacquet13} considered the case of deuterated water and argued the D/H water distribution in Chondrites implies the Schmidt number is smaller than unity.

Snow-lines, where a gas phase tracer (for example H$_2$O, CO$_2$, CO) condenses out of the gas to form ice below some temperature represents a scenario where a sharp concentration gradient can occur. Species such as H$_2$O, CO$_2$ \& CO are relatively abundant, such that the surface layers can be optically thick. Additionally, model degeneracies make it difficult to constrain snow-lines directly (even with optically thin isotopologues). However, if the production/destruction of yet rarer tracers are regulated by the presence or absence of gas phase H$_2$O, CO$_2$ or CO then these rare tracers could be used as proxies to detect snow-lines. For the CO snow-line it is expected that N$_2$H$^+$ and H$_2$CO will only be abundant when CO freezes out \citep{jorgensen04,walsh12,qi13,Qi_science}. Such an expectation has been born out in observations of star-forming cores \citep{friesen10}, and similar results were obtained in the DISCS {\it SMA} survey of several nearby protoplanetary discs \citep{qi13}. Recently, \citet{Qi_science} imaged a hole in N$_2$H$^+$ using {\it ALMA} at a radius of $\sim 30$ AU, co-incident with the expected location of the CO snow-line (based on a free-out temperature of $\sim 17$~K). The sharpness of the inner edge will depend strongly on the strength of the turbulent diffusion, with weaker diffusion resulting in a sharper hole. 

In this letter we demonstrate how the distribution of a tracer species that is only abundant outside a snow-line (being destroyed inside) is strongly dependant on the Schmidt number, and that {\it ALMA} observations could be able to constrain the Schmidt number in protoplanetary discs. 

%In Section~2 we justify and derive a simple 1D analytic model of the distribution of the tracer species. In Section~3 we calculate simulated interferometric observations of the tracer species distributions . In Section~4 we discuss out results and summarise in Section~5. 

%\appendix

%\section{Appendix material}
\section{Disc Model}
We consider a 1D axis-symmetric disc, where the evolution of the gas surface density ($\Sigma_g$)  and the surface density of any (gas-phase) tracer species ($\Sigma_i$) 
%as a function of cylindrical radius ($R$)%
 is given by \citep[e.g.][]{LBP74,clarke88,birnstiel10,owen11,owen14}:
\begin{eqnarray}
\frac{\partial \Sigma_g}{\partial t}&=&\frac{3}{R}\frac{\partial}{\partial R}\left[R^{1/2}\frac{\partial}{\partial R}\left(\nu\Sigma_g R^{1/2}\right)\right]\label{eqn:gas_disc}\\
\frac{\partial \Sigma_i}{\partial t}&+&\frac{1}{R}\frac{\partial}{\partial R}\left[ R\Sigma_i u_g - D^g_R R\Sigma_g\frac{\partial X_i}{\partial R}\right]=\nonumber\\ &&\sum_jS_i(\Sigma_g,\Sigma_j,R,t)\label{eqn:tracer_disc}
\end{eqnarray}
where $X_i$ is the concentration of the tracer species, {\bl $u_g$ is the net radial gas velocity}, $D^g_R$ is the radial gas turbulent diffusion co-efficient and $S_i(\Sigma_g, \Sigma_j,R,t)$ is a source/sink term that represents the production and destruction of the tracer species. 
%In the most general case the source function depends on the surface density of gas, surface density of all other chemical species in the disc, radius and time. 
\subsection{Conditions at the snow-line}
The following general model can be applied to any snow-line (H$_2$O, CO$_2$, CO, etc.) where the chemical and transport time-scales decouple. Here we specifically consider the case of N$_2$H$^+$ destruction at the CO snow-line as observed in TW Hya \citep{Qi_science}. Inside the CO snow-line N$_2$H$^+$ is destroyed by gas phase CO, outside the CO snow line this destruction channel is no-longer dominant and it instead is destroyed at a much slower rate by dissociative recombination \citep{jorgensen04}. Simulations without transport suggest the N$_2$H$^+$ abundance drops by several orders of magnitude inside the CO snow-line \citep[e.g.][]{walsh12} with an abundance that  depends on the square of the gas-phase CO abundance \citep{jorgensen04}.
\subsubsection{Relevant Time-scales}
In order for chemical tracers at the snow-line to be useful in terms of probing the strength of the turbulent diffusion, we must de-couple the chemical and dynamical time-scales. Namely, the desorption time-scale must be faster than the transport time-scales in-order to create a sharp snow-line; furthermore, the destruction time-scale of the tracer species inside the snow-line must also be fast. The transport time-scales of interest are the time with which to move a radial distance $H$ (where $H$ is the disc's scale height\footnote{We note since the snow-line is sharp, the relevant length scale for computing time-scales is $H$, not $R$.} which is of order the radial scale length). Therefore, the advection time-scale ($t_{\rm adv}$) is:
\begin{eqnarray}
t_{\rm adv}&\approx&\frac{H}{u_g}=\frac{2}{3}\alpha^{-1}\left(\frac{R}{H}\right)\Omega^{-1}\nonumber \\ &\approx & 2\times10^{4}\mbox{ years } \left(\frac{\alpha}{0.01}\right)^{-1}\left(\frac{H/R}{0.1}\right)^{-1}\nonumber\\ &&\times \left(\frac{R_{\rm SL}}{30\mbox{ AU}}\right)^{3/2}\left(\frac{M_*}{1\mbox{ M}_\odot}\right)^{-1/2}
\end{eqnarray}
and the diffusive time-scale ($t_{\rm dif}$) is:
\begin{equation}
t_{\rm dif}\approx\frac{RH}{D^g_R}\left(\frac{\partial\log X}{\partial\log R}\right)^{-1}
\end{equation}
It is well known \citep[e.g.][]{clarke88,jacquet12,jacquet13}\footnote{We will derive this dependence in Section~\ref{sec:steady} for a steady disc model, but it is a more general result - see Clarke \& Pringle (1988).} the logarithmic concentration gradient rapidly approaches $3\nu/2D^R_g$, thus the diffusive time-scale is identical to the advection time-scale (this is somewhat unsurprising as they are both driven by the same  process). Thus, the relevant time-scale for the movement of a individual tracer molecule over a radial scale $H$ is ($t_{adv+dif}$) $\sim 10^{4}$ years.
%\begin{eqnarray}
%t_{adv+dif}&=&\left(t_{\rm adv}^{-1}+t_{\rm dif}^{-1}\right)^{-1}\nonumber \\
%&\approx& 1\times10^{4}\mbox{ years } \left(\frac{\alpha}{0.01}\right)^{-1}\left(\frac{H/R}{0.1}\right)^{-1}\left(\frac{R_{\rm SL}}{30\mbox{ AU}}\right)^{3/2}\left(\frac{M_*}{1\mbox{ M}_\odot}\right)^{-1/2}
%\end{eqnarray}

We want to compare this transport time-scale to the time-scale for the desorption of the snow-line species, along with the destruction of  the tracer species inside the snow-line. Considering our example of N$_2$H$^+$ and the CO snow-line then the desorption time-scale is obtained by balancing desorption with absorption (with rate constant $k_S({\rm CO})$), so the desorption time-scale becomes \citep{takahashi00}:
\begin{eqnarray}
t_{\rm dorb} & =& \frac{1}{k_S({\rm CO}) n_{\rm CO}}\nonumber\\
&\approx & 10 \mbox{ years } \mu^{-2} \left(\frac{a}{1\mbox{ mm}}\right)\left(\frac{X_d}{0.01}\right)^{-1}\left(\frac{X_{\rm CO}}{10^{-4}}\right)^{-2}  \nonumber\\&&\times \left(\frac{\Sigma}{1\mbox{ g cm}^{-3}}\right)^{-2} \left(\frac{H/R}{0.1}\right)^2\left(\frac{R}{30\mbox{ AU}}\right)^2
\end{eqnarray}
where $a$ is the dust-grain size, $X_d$ is the dust-to-gas mass ratio and $X_{\rm CO}$ is the CO abundance.  Additionally, the destruction time-scale for N$_2$H$^+$ is (from the \citealt{jorgensen04} simplified network):
\begin{eqnarray}
t_{\rm des}&=&\frac{1}{k_{\rm des}X_{\rm CO}n_g}\nonumber \\
&\approx& 100 \mbox{ years }\mu^{-1} \left(\frac{X_{\rm CO}}{10^{-4}}\right)^{-1}  \left(\frac{\Sigma}{1\mbox{ g cm}^{-3}}\right)^{-1}\nonumber\\&&\times \left(\frac{H/R}{0.1}\right)\left(\frac{R}{30\mbox{ AU}}\right)
\end{eqnarray}
where $k_{\rm des}$ is the destruction rate constant\footnote{Estimated  from Figure~16 of \citet{jorgensen04}.}. 

%The formation of is linked to the cosmic ray ionization rate \citep{jorgensen04}. Thus the formation time-scale is approximately:
%\begin{eqnarray}
%t_{\rm form}&=&X_{\mbox{N$_2$H$^+$}}\zeta^{-1}\nonumber\\
%&\approx& 3\mbox{ years } \left(\frac{X_{\mbox{N$_2$H$^+$}}}{10^{-9}}\right)\left(\frac{\zeta}{10^{-17}\mbox{ s}^{-1}}\right)^{-1}
%\end{eqnarray}
%where $\zeta$ is the cosmic-ray ionization rate and $X_{\mbox{N$_2$H$^+$}}$ is the equilibrium abundance of N$_2$H$^+$, estimated from the observed value in the TW Hya observation \citep{Qi_science}. Thus the time-scale for the snow-line region to reach chemical equilibrium $t_{\rm ch}\sim t_{\rm des}$.

Therefore, the CO/N$_2$H$^+$ system clearly satisfies $t_{\rm des+dorb} \ll t_{\rm adv+dif}$, for  conditions experienced in protoplanetary discs. As such we may ignore the details of the chemical rate equations and simply model the destruction of  any remaining N$_2$H$^+$ to occur instantaneously at the snow-line radius; although, we emphasise that the following analysis can be applied to any tracer species with similar destruction time-scales. Therefore, in this situation the source function $\sum_jS_i(\Sigma_g,\Sigma_j,R,t)$ is drastically simplified to:
\begin{equation}
S_i(R,t)=-\frac{\dot{M}X_i^\infty}{2\pi}\frac{\delta\left(R-R_{\rm SL}(t)\right)}{R}\label{eqn:source}
\end{equation}
where $\dot{M}$ is the mass-accretion rate, $X_i^\infty$ is the concentration at large radius, $\delta(R)$ is the Dirac delta-function and $R_{\rm SL}$ is the radius of the snow-line. This source function represents the instantaneous destruction of any remaining tracer species at $R=R_{\rm SL}$. We note it ignores any possible vertical structure of the snow-line, which will necessarily spread out the destruction region of the tracer species \citep{walsh12} and we discuss the implications of this caveat in Section~4.

\subsection{Steady-disc models}\label{sec:steady}
\begin{figure}
\centering
\includegraphics[width=\columnwidth]{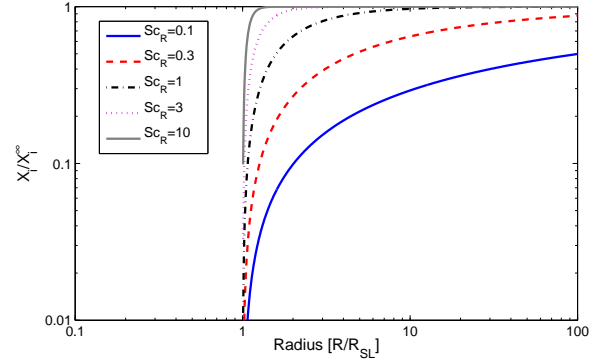}
\caption{The normalised concentration gradient as a function of radius for different values of $Sc_R$ in the range 0.1-10.}\label{fig:conc}
\end{figure}

We will now restrict ourselves to a steady disc problem. In that case Equation~\ref{eqn:gas_disc} becomes $\dot{M}=3\pi\nu\Sigma_g$:
%\begin{equation}
%\nu\Sigma_g=\frac{\dot{M}}{3\pi}\left(1-\sqrt{\frac{R_{\rm in}}{R}}\right)\approx\frac{\dot{M}}{3\pi}\label{eqn:gas_steady}
%\end{equation}
where $\dot{M}=-2\pi R u_g  \Sigma_g$ is the accretion rate, and we have neglected the very small contribution due to an inner boundary at finite radius. Furthermore, the equation for the gas tracer becomes:
\begin{equation}
\frac{\partial}{\partial R}\left[ R\Sigma_i u_g - \frac{\nu}{Sc_R} R\Sigma_g\frac{\partial}{\partial R}\left(\frac{\Sigma_i}{\Sigma_g}\right) \right]=R S_i(R)\label{eqn:tracer_steady}
\end{equation}
where $Sc_R$ is the radial Schmidt number. Therefore, radially integrating Equation~\ref{eqn:tracer_steady} and using our expression for $\dot{M}$ we find:
\begin{equation}
-\frac{\dot{M}}{2\pi}X_i-\frac{\dot{M}}{3\pi Sc_R}R\frac{\partial X_i}{\partial R}=\int^R\!\!\!\!{\rm d}R'\,\, R'S_i(R')\label{eqn:conc_steady_intm}
\end{equation}
%where $X_i$ is the concentration of the tracer species. Equation~\ref{eqn:conc_steady_intm} can be manipulated to read:
%\begin{equation}
%\frac{\partial X_i}{\partial R}+\frac{3Sc_R}{2}\frac{X_i}{R}=\frac{3\pi Sc_R}{\dot{M} R}\int^R\!\!\!\!{\rm d}R'\,\, R'S_i(R')\label{eqn:conc_diff_eqn}
%\end{equation}
Using Equation~\ref{eqn:source} we can integrate Equation~\ref{eqn:conc_steady_intm} to find the radial  concentration distribution:
\begin{equation}
X_iR^{3Sc_R/2}=\int_{0}^R\!\!\!\!{\rm d}R'\frac{3Sc_RX_\infty}{2}{\bl R'}^{3Sc_R/2-1}\Theta({\bl R'}-R_{\rm SL})
\end{equation} 
where $\Theta(R)$ is the Heaviside step function. Setting $X_i=0$ at $R=R_{\rm SL}$ we find the solution:
\begin{equation}
\frac{X_i(R)}{X_i^\infty}=\begin{cases} 1-\left(\frac{R}{R_{\rm SL}}\right)^{-3Sc_R/2} & \mbox{ if } R >R_{\rm SL} \\
0 & \mbox{ if } R\le R_{\rm SL}
\end{cases}\label{eqn:conc}
\end{equation}
 Therefore, we see that the radial profile of the concentration is strongly sensitive to the value of the Schmidt number. In Figure~\ref{fig:conc} we show how the concentration  varies with radius and Schmidt number. It is important to emphasise that concentration distribution is independent of assumptions of the (unknown) viscosity, and mass-accretion rate and can in principle provide a `clean' measurement of the Schmidt number. 

\section{Observable characteristics}
Unfortunately it is not possible to directly observe the concentration gradient. What is directly observed is the surface-brightness distribution of the relevant species. The surface-brightness distribution is sensitive to the surface density of the tracer species rather than its concentration. Thus, we must multiply the concentration of our tracer species by the gas surface density. Adopting a power-law gas distribution of the form $\Sigma_g/\Sigma_{\rm SL}= (R/R_{\rm SL})^{-\gamma}$ with a cut-off radius $R_{\rm out}$, the surface density of the tracer species is:
%\begin{equation}
%\Sigma_{t}=\begin{cases} 0 & \mbox{ if } R\le R_{\rm SL} \\
%X_{\infty}\Sigma_{SL}\left[\left(\frac{R}{R_{\rm SL}}\right)^{-\gamma  } -\left(\frac{R}{R_{\rm SL}}\right)^{-3/2Sc_R-\gamma}\right] & \mbox{ if } R_{\rm SL}<R<R_{\rm out} \\
%0 & \mbox{ if } R\ge R_{\rm out}
%\end{cases}\label{eqn:tracer_surf}
%\end{equation}
\begin{equation}
\Sigma_{t}=X_{\infty}\Sigma_{SL}\left[\left(\frac{R}{R_{\rm SL}}\right)^{-\gamma  } -\left(\frac{R}{R_{\rm SL}}\right)^{-3/2Sc_R-\gamma}\right] \label{eqn:tracer_surf}
\end{equation}
in the range $R_{\rm SL}<R<R_{\rm out}$ and $\Sigma_t=0$ elsewhere.
\subsection{Brightness distribution and Visibility profiles}
The actual brightness distribution from rotational emission lines (such as N$_2$H$^+$ J=4-3), as well as being sensitive to the surface density distribution is also sensitive to the background gas temperature and excitation temperature of the molecule (which in turn is a function of density and temperature) along with the optical depth. 

However, in the case that the tracer species is optically thin, and the gas density is far above the critical density ($n_{cr}$) for the rotational transition, then the molecular line is thermalised. Thus, if the temperature gradient is weak compared to the scales of interest and $n_g\gg n_{cr}$ then we can approximate the surface brightness as being directly proportional to the surface density of the tracer species, provided the line remains optically thin. 

For the gas at 30~AU the gas density is typically:
\begin{eqnarray}
n_g&=&5.3\times10^{9}\mbox{ cm}^{-3}\,\mu^{-1} \left(\frac{\Sigma}{1\mbox{ g cm}^{-3}}\right)\nonumber\\ &&\times \left(\frac{H/R}{0.1}\right)^{-1}\left(\frac{R}{30\mbox{ AU}}\right)^{-1}
\end{eqnarray}
where $\mu$ is the mean molecular weight of the gas. Comparing this to the critical density of the $J=4-3$ transition of N$_2$H$^+$ which has a critical density of $n_{cr}\sim 10^{7}$ cm$^{-3}$ \citep[e.g.][]{friesen10}, we see that $n_g\gg n_{cr}$. We note, since temperature and density are expected to be power-laws with radius \citep[e.g.][]{chiang97,hartmann98} then the small additional corrections due to  temperature and density effects of converting $\Sigma_i$ to the surface brightness distribution ($B_t$) will manifest themselves as changes in the power-law index $\gamma$ in Equation~\ref{eqn:tracer_brightness}. 

Assuming the disc to be axis-symmetric and observed face-on, we may write the brightness distribution on the sky as:
%\begin{equation}
%B_t(\theta)\approx\begin{cases}0 & \mbox{ if } \theta\le \theta_{\rm SL} \\
%B_0\left[\left(\frac{\theta}{\theta_{\rm SL}}\right)^{-\gamma  } -\left(\frac{\theta}{\theta_{\rm SL}}\right)^{-3/2Sc_R-\gamma}\right] & \mbox{ if } \theta_{\rm SL}<\theta<\theta_{\rm out} \\
%0 & \mbox{ if } \theta\ge \theta_{\rm out}
%\end{cases}\label{eqn:tracer_brightness}
%\end{equation}
\begin{equation}
B_t(\theta)\approx B_0\left[\left(\frac{\theta}{\theta_{\rm SL}}\right)^{-\gamma  } -\left(\frac{\theta}{\theta_{\rm SL}}\right)^{-3/2Sc_R-\gamma}\right] \label{eqn:tracer_brightness}
\end{equation}
in the range $\theta_{\rm SL}{\bl <} \theta<\theta_{\rm out}$ and $B_t=0$ elsewhere, where $B_0$ is a constant, $\theta$ is the angular size on the sky and $\theta_{SL}$ is the angular size of the snow-line given by:
\begin{equation}
\theta_{SL}=0.2 \mbox{ arcsec} \left(\frac{R_{SL}}{\mbox{30 AU}}\right)\left(\frac{d}{\mbox{150 pc}}\right)^{-1}
\end{equation}

Thus, we see that the brightness distribution still retains the strong sensitivity to the Schmidt number. Since the angular resolution required to probe the value of the Schmidt number is only available through mm-interferometry, we can use our brightness distribution to calculate synthetic visibilities.
%Assuming the source to be axis-symmetric and fitting the (real) visibilities to the axis-symmetric model will provide the highest sensitivities. 
The visibilities can obtained by a Hankel transform of the axis-symmetric brightness-distribution such that:
\begin{equation}
V_t(\eta)=2\pi\int_0^\infty\!\!\!\!{\rm d}\theta\theta B_t(\theta)J_0(2\pi\eta\theta)
\end{equation}
where $\eta$ is a radial baseline co-ordinate defined as $\eta=\sqrt{u^2+v^2}$ where $u$ \& $v$ are the usual baseline co-ordinates. Following the observed N$_2$H$^+$ $J=4-3$ emission by \citet{Qi_science} we calculate our synthetic observations at 372Ghz with our standard model adopting the best fit parameters from \citet{Qi_science} of $R_{\rm SL}=30$ AU, $R_{\rm out}=150$ AU and $\gamma=2$. 
%Assuming a snow-line radius of $30$ AU and $R_{\rm out}$=200 AU we calculate the expected Visibility curves for the snow-line tracer species (e.g. N$_2$H$^+$) for different values of the Schmidt number and power-law index of the gas surface density ($\gamma$) are shown in Figure~\ref{fig:brightness_vis}. In the top panels we show the visibility curves for a source at 50pc (comparable to TW Hya) and in the bottom panel we show the visibility curves for a source at 150pc (comparable to objects in Gould Belt star forming regions). In the left panels we vary the Schmidt number for fixed $\gamma=1$ and in the right panels we vary $\gamma$ for a fixed Schmidt number of $Sc_R=1$.

\begin{figure*}
\centering
\includegraphics[width=\textwidth]{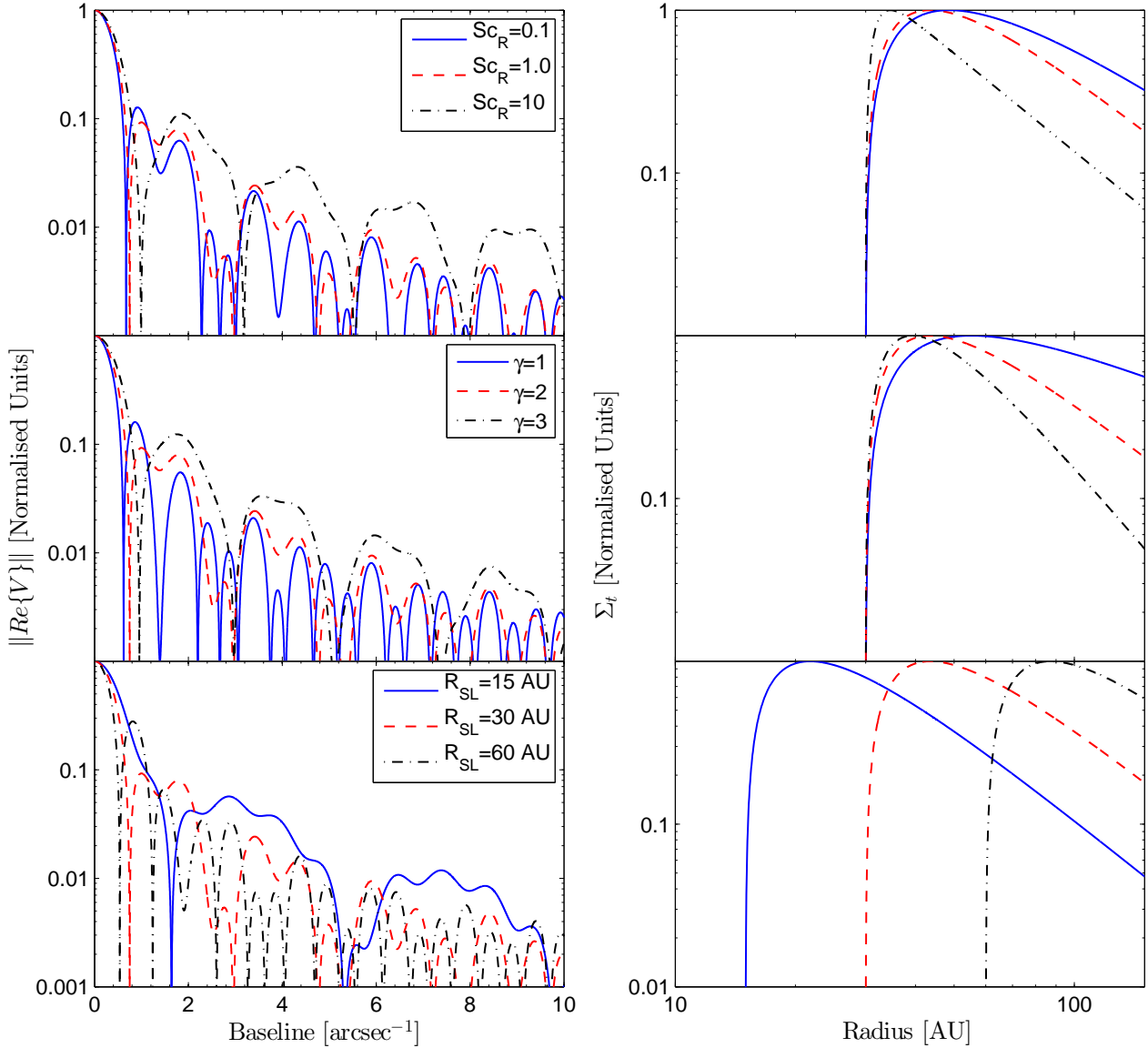}
\caption{Predicted visibility profiles assuming a source distance of 150~pc (left panels) and tracer surface density profiles. In the top panel we vary the Schmidt number between 0.1-10; in the middle panels we vary the background gas profile exponent ($\gamma$) between 1-3; and in the bottom panels we vary the snow-line radius between 15-60 AU. The visibilities are normalised to 1 at $\xi=0$ and the surface density profiles are normalised so that the peak value has $\Sigma=1$. }\label{fig:brightness_vis}
\end{figure*}

We calculate our synthetic observations assuming the source is observed face-on at a distance of 150pc. In Figure~\ref{fig:brightness_vis} we show our simulated visibility curves in the left-hand panels and the surface density profiles in the right hand panel. In the top panels we vary the Schmidt number between 0.1-10, in the middle panels we vary $\gamma$ from 1-3 and in the bottom panels we vary $R_{\rm SL}$ from 15-60 AU. 

The visibility curves in Figure~\ref{fig:brightness_vis} clearly show that variations in the Schmidt number give rise to significant differences. Furthermore, comparisons between varying the index of the gas surface density $\gamma$, snow-line radius and the Schmidt number are not too degenerate. The simplified model presented here contains four free parameters $\{R_{\rm SL},R_{\rm out}, Sc_R, \gamma\}$, which would all need to be constrained by fitting the visibilities. Inspection of Figure~\ref{fig:brightness_vis} suggests that sensitivities $\lesssim 10$ \% at base-lines $\sim 0.1-1$ km would allow the theoretically suggested range of Schmidt numbers (0.1-10) to be observationally constrained. %The high sensitivity and large number of available base-lines with {\it ALMA} suggests that {\it ALMA} observations of snow-line proxies in proto-planetary discs will be able to constrain the possible value of the Schmidt number in the outer regions of protoplanetary discs.    

\section{Discussion}
We have shown that observations of snow-lines in protoplanetary discs using  a tracer species (for example N$_2$H$^+$ in the case of the CO snow line \citealt{qi13,Qi_science}) can be used to probe the Schmidt number, a unknown parameter in studies of turbulent transport in accretion discs, where current estimates span a range of two-orders of magnitude ($Sc_R=$ 0.1-10). 
%and we have indicated that {\it ALMA} observations could constrain the value significantly. 

\subsection{Detectability with {\it ALMA}}

The {\it ALMA} telescope is a sub-mm/mm interferometer; once completed it will have $\sim 50$ individual antennas, offering $\gtrsim 1000$ baselines with separations upto 12~km. At 372 Ghz ($\sim$ N$_2$H$^+$ J=4-3 line) this provides a maximum spatial resolution of $\sim 0.01$ arcsec. Figure~\ref{fig:brightness_vis} clearly shows that {\it ALMA} posses the required number of base-lines ($\gtrsim 50$) with separations in the range 0.1-1 Km to constrain Schmidt values within the current range of uncertainty, provided the observations are sensitive enough.  Taking the TW Hya N$_2$H$^+$ observations as reference \citep{Qi_science} - a source brightness $\sim 200$ mJy beam$^{-1}$ km s$^{-1}$ with an rms noise of 8.1 mJy beam$^{-1}$ km s$^{-1}$ and beam size $\sim 0.6$ arcsec made using 23-26 antennas\footnote{\bl Since the beam size of the current TW Hya observation posses a resolution similar to $\theta_{\rm SL}$ then using the current observation to constrain the Schmidt number seems unlikely; however, using the velocity channels separately can be used to increase the effective resolution \citep[e.g.][]{qi13}.}   - a similar level of sensitivity,  but with a beam size of $\lesssim 0.3$ arcsec, could be obtained assuming a fully operational {\it ALMA} of $\sim$50 antennas. Thus, such observations  are feasible and sufficiently high resolution to allow constraints to be placed on the Schmidt number at the distance to TW Hya. Longer integration times and larger base-lines would be required to reach similar levels of sensitivity at distances of 150pc.          

%Fully operational {\it ALMA} sensitivity of 0.01mJy in a 10~hour integration.

\subsection{Uncovering properties of the turbulence}
We have argued that the sharpness of the hole in tracer species at the snow-line probes the value of the Schmidt number independent of the assumed properties of the turbulence (e.g. assumed value of the viscous `$\alpha$' parameter). Since several snow-lines are expected to occur at different radii, then measurements at different snow-lines would allow the radial dependence of the Schmidt number to be probed. In particular, recent simulation suggest that different non-ideal MHD effects (which dominate at different radii \citealt{turner14}) lead to different Schmidt numbers \citep{zhu14}. {\bl Thus, comparing the simulation predictions of the Schmidt number for various turbulent driving mechanisms with the observed value would allow inferences about the nature of the turbulence to be made.}

{\bl Furthermore, independently measuring the rms turbulent velocity ($\langle v_R^2\rangle$) \citep[e.g.][]{hughes11} then combining it  with a measurement of the Schmidt number would allow the viscosity (including estimates of $\alpha~\mbox{where}~\nu=\alpha H^2\Omega$) to be calculated, since $D_R^g=\alpha H^2\Omega/Sc_R\approx\langle v_R^2\rangle/\Omega$}.

%Thus, {\it ALMA} can begin to place direct observational constrains on the type and strength of turbulent angular momentum and turbulent diffusion in astrophysical accretion discs. 

\subsection{Caveats \& Limitations}
We have constructed an idealised model to investigate whether snow-lines could begin to probe the strength of turbulent diffusion. As such there are several model improvements that must be made before fitting to real data. Therefore, our model presented in this letter is a `proof-of-concept' rather than a road map for observational modelling. For example, a real protoplanetary disc is not one-dimensional. As such the vertical temperature structure is not constant and passively heated discs cool as one approaches the mid-plane \citep{chiang97}. Therefore, the snow-line is unlikely to occur exactly at a fixed radius, but is more-likely to be an extended structure with a scale variation of $\sim H$, with is time-varying position \citep{martin12,martin13,martin14}. Additionally the conversion of the gas phase to ice particles at the snow-line will result in turbulent diffusion of the gas and ice particles away from the snow-line (in identical manner to that discussed for the snow-line tracer discussed here). As such, there is unlikely to be a very sharp change in the gas abundance at the snow-line but rather a smoother change. Furthermore, the chemical time-scales may not fully decouple from the transport time-scales. In the N$_2$H$^+$ case considered here we have argued that the time-scales are likely to be decoupled; this may not be the case for all snow-line tracer species, thus dynamical modelling which includes turbulent diffusion is needed to determine the importance of this effect.  {\bl The model presented here should provide a stringent upper limit of the Schmidt number, and good measurement if it is small ($<1$); however, if it is large, and the sharpness of the passive tracer has width of $\sim H$, then the 1D model would only provide an order of magnitude estimate and a better model is need to constrain the Schmidt number.} {\bl Finally, if the snow-line resides in a dead-zone, where there is limited or no turbulence then it is unlikely this method can be used to cleanly probe the Schmidt number; but, dead-zones are not expected at the large radius of the CO/N$_2$H$^+$ system discussed here.}

%All the additional considerations (vertical structure, turbulent diffusion of the ice/gas species and chemical time-scales) will result in our model over-estimating the sharpness of the hole in the tracer species. Therefore, any measurement of the Schmidt number will provide an upper limit. We note many of these additional considerations could be modelled for. For example a 2D model that accounts for vertical diffusion (through a vertical Schmidt number $Sc_z$), coupled with observations of discs observed at various inclinations would allow the ratio of $Sc_z/Sc_R$ to be probed, a ratio which is strongly dependant on non-ideal MHD effects \citep{zhu14}.  

\section{Summary}
In this letter we have shown that the recent observations of snow-lines through tracer species (e.g. N$_2$H$^+$ or H$_2$CO in the case of the CO snow-line \citealt{qi13,Qi_science}) could allow direct observational measurements of the Schmidt number in astrophysical accretion discs. In the case that the chemical time-scale is suitably de-coupled from the transport time-scales then the concentration gradient of the tracer outside the snow-line directly depends on the Schmidt number in a power-law fashion ($\sim R^{-3/2Sc_R}$) independent of the choice of the turbulent $\alpha$ parameter.

We argue that the effect of turbulent diffusion on surface brightness distribution of such a snow line tracer is detectable with {\it ALMA} observations of discs in nearby star-forming regions which can possess high enough angular resolution to constrain the current theoretically/numerically estimated values of the Schmidt number ($Sc_R\sim 0.1-10$).   

Observations of different snow-lines (e.g. H$_2$O, CO$_2$ \& CO) at different radii in the discs would allow the radial dependence of the Schmidt number to be probed. Coupling these snow-line observations with observational estimates of the gas-surface density and turbulent line-widths would allow direct estimates of the strength and nature of the turbulence in astrophysical accretion discs.

\acknowledgements
We thank the anonymous referee for a helpful comments on the manuscript. JEO is grateful to Karin \"Oberg for a discussion that sparked this investigation and to Rachel Friesen for helpful advice.

\bibliographystyle{apj}
%\bibliography{bib_paper}

%\begin{thebibliography}
%\item{hello} Hello
%\end{thebibliography}

\end{document}